\documentclass[twocolumn]{aastex631}
\usepackage{amsmath} 

\received{June 01, 2024}
\revised{September 24, 2024}
\accepted{October 13, 2024}

\shortauthors{Su et al.}

\begin{document}

\title{Could the inter-band lag of active galactic nucleus vary randomly?}

\email{zbsu@mail.ustc.edu.cn; zcai@ustc.edu.cn}

\author[0000-0001-8515-7338]{Zhen-Bo Su}
\affiliation{Department of Astronomy, University of Science and Technology of China, Hefei 230026, China}
\affiliation{School of Astronomy and Space Science, University of Science and Technology of China, Hefei 230026, China}

\author[0000-0002-4223-2198]{Zhen-Yi Cai}
\affiliation{Department of Astronomy, University of Science and Technology of China, Hefei 230026, China}
\affiliation{School of Astronomy and Space Science, University of Science and Technology of China, Hefei 230026, China}

\author[0000-0002-4419-6434]{Jun-Xian Wang}
\affiliation{Department of Astronomy, University of Science and Technology of China, Hefei 230026, China}
\affiliation{School of Astronomy and Space Science, University of Science and Technology of China, Hefei 230026, China}

\author[0000-0002-1517-6792]{Tinggui Wang}
\affiliation{Department of Astronomy, University of Science and Technology of China, Hefei 230026, China}
\affiliation{School of Astronomy and Space Science, University of Science and Technology of China, Hefei 230026, China}

\author[0000-0002-1935-8104]{Yongquan Xue}
\affiliation{Department of Astronomy, University of Science and Technology of China, Hefei 230026, China}
\affiliation{School of Astronomy and Space Science, University of Science and Technology of China, Hefei 230026, China}

\author[0000-0003-4721-6477]{Min-Xuan Cai}
\affiliation{Department of Astronomy, University of Science and Technology of China, Hefei 230026, China}
\affiliation{School of Astronomy and Space Science, University of Science and Technology of China, Hefei 230026, China}

\author[0000-0003-4200-4432]{Lulu Fan}
\affiliation{Department of Astronomy, University of Science and Technology of China, Hefei 230026, China}
\affiliation{School of Astronomy and Space Science, University of Science and Technology of China, Hefei 230026, China}
\affiliation{Deep Space Exploration Laboratory, Hefei 230088, China}

\author[0000-0001-8416-7059]{Hengxiao Guo}
\affiliation{Shanghai Astronomical Observatory, Chinese Academy of Sciences, 80 Nandan Road, Shanghai 200030, China}

\author[0000-0003-3667-1060]{Zhicheng He}
\affiliation{Department of Astronomy, University of Science and Technology of China, Hefei 230026, China}
\affiliation{School of Astronomy and Space Science, University of Science and Technology of China, Hefei 230026, China}

\author[0000-0001-8554-9163]{Zizhao He}
\affiliation{Purple Mountain Observatory, Chinese Academy of Sciences, Nanjing 210023, China}

\author[0009-0006-9884-6128]{Xu-Fan Hu}
\affiliation{Tsung-Dao Lee Institute, Shanghai Jiao-Tong University, Shanghai, 520 Shengrong Road, 201210, People's Republic of China}

\author[0000-0002-9092-0593]{Ji-an Jiang}
\affiliation{Department of Astronomy, University of Science and Technology of China, Hefei 230026, China}
\affiliation{School of Astronomy and Space Science, University of Science and Technology of China, Hefei 230026, China}

\author[0000-0002-7152-3621]{Ning Jiang}
\affiliation{Department of Astronomy, University of Science and Technology of China, Hefei 230026, China}
\affiliation{School of Astronomy and Space Science, University of Science and Technology of China, Hefei 230026, China}

\author[0000-0003-2573-8100]{Wen-Yong Kang}
\affiliation{Department of Astronomy, University of Science and Technology of China, Hefei 230026, China}
\affiliation{School of Astronomy and Space Science, University of Science and Technology of China, Hefei 230026, China}

\author[0000-0003-4631-1915]{Lei Lei}
\affiliation{Purple Mountain Observatory, Chinese Academy of Sciences, Nanjing 210023, China}
\affiliation{School of Astronomy and Space Science, University of Science and Technology of China, Hefei 230026, China}

\author[0000-0003-4286-5187]{Guilin Liu}
\affiliation{Department of Astronomy, University of Science and Technology of China, Hefei 230026, China}
\affiliation{School of Astronomy and Space Science, University of Science and Technology of China, Hefei 230026, China}

\author[0000-0002-2941-6734]{Teng Liu}
\affiliation{Department of Astronomy, University of Science and Technology of China, Hefei 230026, China}
\affiliation{School of Astronomy and Space Science, University of Science and Technology of China, Hefei 230026, China}

\author[0000-0002-2242-1514]{Zhengyan Liu}
\affiliation{Department of Astronomy, University of Science and Technology of China, Hefei 230026, China}
\affiliation{School of Astronomy and Space Science, University of Science and Technology of China, Hefei 230026, China}

\author[0000-0001-6938-8670]{Zhenfeng Sheng}
\affiliation{Deep Space Exploration Laboratory, Hefei 230088, China}

\author[0000-0002-0771-2153]{Mouyuan Sun}
\affiliation{Department of Astronomy, Xiamen University, Xiamen 361005, China}

\author[0000-0002-1330-2329]{Wen Zhao}
\affiliation{Department of Astronomy, University of Science and Technology of China, Hefei 230026, China}
\affiliation{School of Astronomy and Space Science, University of Science and Technology of China, Hefei 230026, China}



\begin{abstract}

The inter-band lags among the optical broad-band continua of active galactic nuclei (AGNs) have been intensively explored over the past decade. However, the nature of the lags remains under debate. Here utilizing two distinct scenarios for AGN variability, i.e., the thermal fluctuation of accretion disk and the reprocessing of both the accretion disk and clouds in the broad line region, we show that, owing to the random nature of AGN variability, the inter-band lags of an individual AGN would vary from one campaign with a finite baseline to another. Specifically, the thermal fluctuation scenario implies larger variations in the lags than the reprocessing scenario. Moreover, the former predicts a positive correlation between the lag and variation amplitude, while the latter does not result in such a correlation. For both scenarios, averaging the lags of an individual AGN measured with repeated and non-overlapping campaigns would give rise to a stable lag, which is larger for a longer baseline and gets saturation for a sufficiently long baseline. However, obtaining the stable lag for an individual AGN is very time-consuming. Alternatively, it can be equivalently inferred by averaging the lags of a sample of AGNs with similar physical properties, thus can be properly compared with predictions of AGN models. In addition, discussed are several new observational tests suggested by our simulations as well as the role of the deep high-cadence surveys of the Wide Field Survey Telescope in enriching our knowledge of the lags.

\end{abstract}



\section{Introduction} \label{sec:intro}

Active galactic nuclei (AGNs) have long been postulated to be powered by the process of gas accretion into a supermassive black hole (BH), resulting in the formation of an accretion disk and producing enormous energy radiations over the whole electromagnetic spectrum. Hitherto, the most widely adopted model for the accretion disk in AGNs is the optically thick and geometrically thin disk model (\citealt{ShakuraSunyaev1973A}; hereafter the SSD model), though its validation has long been questioned \citep[e.g.,][]{Antonucci2013Natur.495..165A,Antonucci2015arXiv150102001A,Antonucci2018NatAs...2..504A,Antonucci2023Galax..11..102A,Cai2023NatAs.tmp..212C}.

Notwithstanding the fact that the accretion disks in almost all AGNs are currently inaccessible to direct imaging, two alternative methods have been proposed to infer the disk size. One probe is the microlensing, by which the emission from a smaller disk region of a background quasar (i.e., the very luminous AGN) could be more significantly fluctuated by stars in the foreground lensing galaxy. Up to now, the microlensing probe has been utilized in only tens of gravitationally lensed quasars \citep[e.g.,][]{Jimenez-Vicente2014ApJ...783...47J,Bate2018MNRAS.479.4796B,Cornachione2020ApJ...895...93C}, and the inferred disk sizes in optical are generally found to be larger than the SSD prediction by $\sim$ 0.5 dex on average \citep[e.g.,][]{Dai2010, Morgan2018}.

Another probe is the so-called continuum reverberation mapping (CRM), in which the UV/optical continuum variation is assumed to be driven by heating of the fluctuating X-ray emission from the vicinity of the central BH, so the variation at shorter wavelength from the inner disk leads that at longer wavelength from the outer disk \citep{Krolik1991}. This assumption is usually designated as the X-ray reprocessing.

Compared to the microlensing method, the CRM method is more attractive since it could be applied to a large number of non-lensed normal AGNs and quasars whose multi-band light curves (LCs) are easily obtained in the time domain era. Now, disk sizes of several hundreds of normal AGNs and quasars have been estimated using the CRM method, and hundreds of thousands more would be available thanks to several ongoing and upcoming high-cadence multi-wavelength surveys \citep[e.g.,][]{Brandt2018,wfst2023}.

A small number of AGNs and quasars have been well monitored in multiple UV/optical bands, either in daily cadence within one year, e.g., the AGN Space Telescope and Optical Reverberation Mapping project \citep[STORM;][]{DeRosa2015ApJ...806..128D} and alike successors, or in sparser cadence ($\gtrsim 3 - 7$ days) but over several years, e.g., the Zwicky Transient Facility survey \citep[ZTF;][]{Graham2019}, the Pan-STARRS1 survey \citep[PS1;][]{Chambers2016}, the Sloan Digital Sky Survey Reverberation Mapping project \citep[SDSS-RM;][]{Shen2015ApJS..216....4S}, and the Dark Energy Survey \citep[DES;][]{Abbott2018ApJS..239...18A}.
However, conflicting results have been obtained when applying the CRM method to these observations. 
The disk sizes of some AGNs and quasars are found to be larger than the SSD prediction by a factor of $\sim 2 - 10$ \citep[e.g.,][]{Fau2016,Jiang2017,Pal2017MNRAS.466.1777P,Fausnaugh2018,Cackett2018,Edelson2019,Jha2022,guo2022,GuoH2022,Montano2022,Kara2023ApJ...947...62K}, while some others are consistent with or just slightly larger than the SSD prediction \citep[e.g.,][]{Kokubo2018,Mudd2018,Homayouni2019,Yu2020,Santisteban2020,Jha2022}.
Meanwhile, the disk size discrepancy appears less prominent for more luminous \citep{Li2021} and massive \citep{GuoH2022} AGNs.

The origin of the disk size discrepancy between observed and SSD-predicted is still unclear, but many solutions to the discrepancy have been proposed, including the inhomogeneous disk fluctuation \citep{DexterAgol2011}, the internal reddening of AGN host galaxy \citep{Gaskell2017MNRAS.467..226G}, the departure from non-blackbody emission \citep{Hall2018ApJ...854...93H}, the role of disk wind \citep{Sun2019MNRAS,Li2019MNRAS.483.2275L}, and the diffuse continuum emission (DCE) from the broad line region \citep[BLR;][]{Cackett2018,Chelouche2019,Netzer2022,Montano2022}. 
All of them are based on the reprocessing scenario and the origin of the inter-band lag is the differential light traveling. Actually, a new origin for the inter-band lag as a result of the differential regression capability of local thermal fluctuation (or the differential capability of re-establishment of local thermal equilibrium after perturbed) has been proposed \citep{Cai2018,Cai2020,Sun2020ApJ...891..178S}. According to the thermal fluctuation scenario, the measured inter-band lags are no longer simply related to the light traveling time and thus the disk size. In other words, the origin of the lags should be clearly understood before utilizing the lag to estimate the disk size. Here a relevant question to be resolved by upcoming surveys is whether the lags measured in different periods are stable.

\begin{figure*}
\centering
\includegraphics[width=0.4\textwidth]{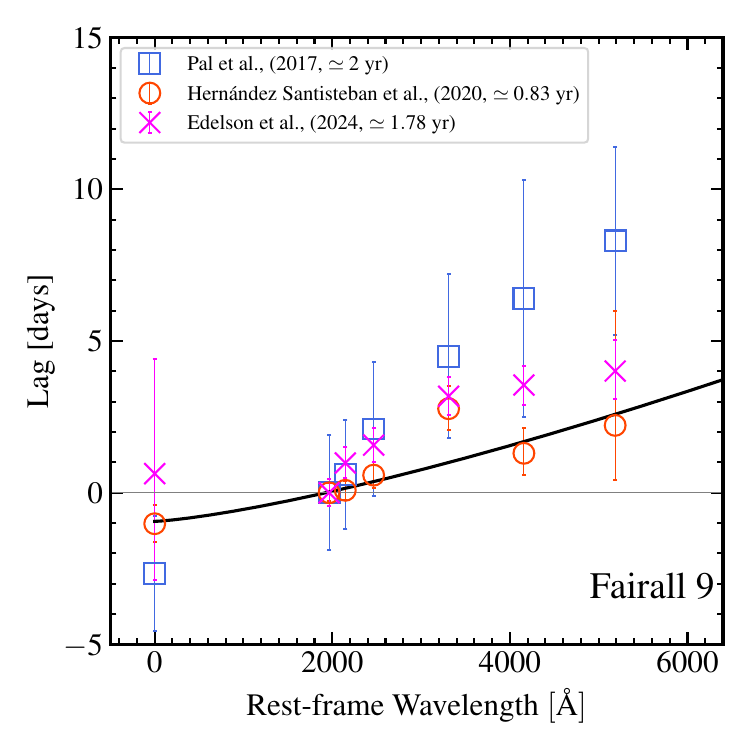}
\includegraphics[width=0.4\textwidth]{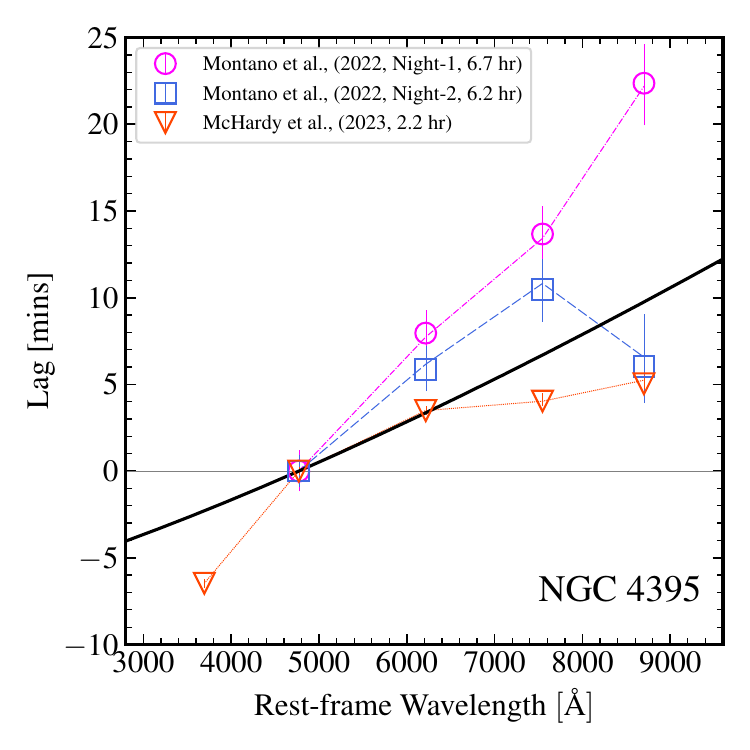}
\caption{The potentially varied lag-wavelength relations observed in two AGNs.
Left panel: relative to the {\it Swift}-$\rm UVW2$ band at 2055 $
\text{\AA}$, the lag-wavelength relations for Fairall 9 are measured in three distinct periods by \citet[][from 2013 April to 2015 April]{Pal2017MNRAS.466.1777P},  \citet[][from 2018 May to 2019 February]{Santisteban2020} and \citet[][from 2018 May to 2020 February]{Edelson2024}. Only the {\it Swift} bands are shown as \cite{Edelson2024}. The black solid line indicates the SSD-predicted lag-wavelength relation for Fairall 9 using a BH mass of $2.6\times10^8M_\odot$ and an Eddington ratio of 0.02 \citep{Vasudevan2009}.
Right panel: relative to the $g$ band, the lag-wavelength relations for NGC 4395 are measured in three nights by \citet[][2.2 hr in 2018 April]{McHardy2023MNRAS.519.3366M} and \citet[][6.7 and 6.2 hr in 2022 April 26 and 27, respectively]{Montano2022}.
The black solid line is the SSD-predicted lag-wavelength relation implied by a BH mass of $1.7\times10^4~M_\odot$ (\citealt{Cho2021ApJ...921...98C}) and a bolometric luminosity of $5.3\times10^{40}~{\rm erg~s^{-1}}$ \citep{Moran2005}. Here the SSD-predicted lag-wavelength relations are simply given by Equation (12) of \cite{Fau2016} assuming $\eta=0.1$, $\kappa = 1$, and $X=5.04$ \citep{tie2018}.}
\label{fig:NGC4395_lag_test}
\end{figure*}

Even good stability of the lags inferred from repeated multi-year campaigns has been reported for a few AGNs, e.g., Mrk 279 \citep{Chelouche2019} and Mrk 110 \citep{Vincentelli2022},
it is highly intriguing that potentially varied lags have been unveiled in some AGNs. First, in a radio-quiet Seyfert type 1 AGN, Fairall 9 (the left panel of Figure~\ref{fig:NGC4395_lag_test}), \citet{Pal2017MNRAS.466.1777P} found very large lags, inconsistent with the SSD prediction, using observations from 2013 April to 2015 April, while \citet{Santisteban2020} showed small lags utilizing observations from 2018 May to 2019 February.  
Further extending the baseline (from 2018 May to 2020 February) for Fairall 9, \citet[][]{Edelson2024} recently reported lags larger\footnote{However, \cite{Edelson2024} suggested the lags are quite stable among four  periods of $\sim 160$ days.} than \citet{Santisteban2020}, albeit with substantial errors. Second, a similar result has been unveiled in the dwarf Seyfert 1 galaxy NGC 4395 (the right panel of Figure~\ref{fig:NGC4395_lag_test}).
Although the lags of the $r$ ($i$ or $z$) band relative to the $g$ band between two successive nights of \citet[][on 2022 April 26 and 27]{Montano2022} differ somewhat at $\simeq 1.1\sigma$ ($\simeq 1.3\sigma$ or $\simeq 4.5\sigma$) confidence level, the differences\footnote{The differences in the mean flux densities between \citet{McHardy2023MNRAS.519.3366M} and \citet{Montano2022} are smaller than 17\% and thus can not be responsible for the lag difference.} become very significant at $\simeq 3.2\sigma$ ($\simeq 5.8\sigma$ or $\simeq 7.0\sigma$) confidence level once comparing the lags measured by \citet[][the first night]{Montano2022} and by \citet[][2018 April 16]{McHardy2023MNRAS.519.3366M}.
Third, \citet[][see their Figure~3]{Vincentelli2023} reported a significant lag change between two $\sim$two-month periods for NGC 7469.

Although the change of the lags could be attributed to distinct observational conditions met by various studies (e.g., different photometric uncertainty and sampling cadence), finite realizations of a stochastic process \citep{Welsh1999}, unresolved problems in data reduction \citep{McHardy2023MNRAS.519.3366M}, or improper methodologies adopted \citep{Chan2020}, does the aforementioned potential change of the lags in a few AGNs hint a random nature for the lags among UV/optical continua? Actually, the lag randomness has been predicted by the thermal fluctuation scenario \citep{Cai2018,Cai2020,Sun2020ApJ...891..178S}, while the conventional reprocessing over an SSD disk gives rise to quite stable lags \citep[][]{Sun2019MNRAS,Vincentelli2022}. Note that a larger scatter in the lags predicted by the reprocessing scenario is possible once considering either a dynamical driving source, e.g., the dynamical evolution of the corona height \citep{Kammoun2021}, or a variable contribution of the BLR emission \citep{Vincentelli2023, Jaiswal2023}.

In this work, being different from the general simulations in the framework of reprocessing \citep[e.g.,][]{Yu2020, Chan2020, Kova2021,Kovacevic2022,pozo2023} for the Deep Drilling Fields (DDF) of the Vera C. Rubin Observatory Legacy Survey of Space and Time \citep[LSST;][]{Brandt2018}, we also utilize the up-to-date thermal fluctuation model (Section~\ref{sec:accretion_disk}) to quantitatively predict the lag randomness using simulations for AGNs in analog to the famous Seyfert 1 galaxy NGC 5548, and present distinguishable properties on the lags predicted by the two distinct scenarios for AGN variability (Section~\ref{sect:predit_lag}).
Next in Section~\ref{WFST} we anticipate how surveys to be conducted by the 2.5-meter Wide Field Survey Telescope \citep[WFST equipped with five SDSS-like $ugriz$ filters;][]{wfst2023} can help improve our understanding on the multi-wavelength AGN variability and the associated inter-band lags. Finally, brief conclusions are presented in Section~\ref{sect:conclusion}.

\section{Models for the optical variability of AGN}\label{sec:accretion_disk}

There are several models proposed to account for the optical multi-band variation properties of AGNs, such as NGC 5548 in particular, which has been intensively monitored for half a year from 2014 January to July \citep{Fau2016}. In other words, models are generally constrained by such observations with limited lengths. However, the constrained models are helpful in not only predicting the variation properties at timescales longer than observed but also providing a way of taking the effect of the randomness of AGN variability into account. 

Here we adopt two distinct models: the thermal fluctuation model (Section~\ref{sect:thermal_fluctuation_model}; \citealt{Cai2018,Cai2020}) and the reprocessing model with DCE, including emissions from both the accretion disk and clouds in the BLR (Section~\ref{sect:reprocessing_model}; \citealt{Cackett2022ApJ,Jaiswal2023}). Utilizing 200 different realizations of 10-year-long WFST-$ugriz$ LCs implied by both models, we will show that dispersions of the lag-wavelength relations predicted by them are very different and thus can be used to distinguish models for AGN variability.

\subsection{Thermal fluctuation model}\label{sect:thermal_fluctuation_model}

Previously, \citet{Cai2018,Cai2020} developed a thermal disk model, exploring the UV/optical continuum lag in AGNs (EUCLIA), in which the UV/optical/X-ray variations are all attributed to disk/corona turbulence, contrary to the reprocessing scenario where the UV/optical variations are due to the varying X-ray/EUV heating.
Actually, there are some known challenges to the reprocessing scenario, including
1) a possible deficit of X-ray energy budget \citep[e.g.,][]{Gaskell2007}; 
2) sometimes poor coordination observed between X-ray and UV/optical variations \citep{Maoz2002,Gaskell2006,Gallo2018,Starkey2017,Morales2019,SouH2022}; 
3) too much high-frequency power predicted \citep{Gardner2017};
4) too small UV/optical variation amplitudes predicted \citep{SouH2022};
5) too small time lags predicted (see references in Section~\ref{sec:intro}); 
and 6) too weak timescale dependence of the color variation predicted \citep{Zhu2018}.

Serving as a choice to circumvent/overcome these challenges encountered by the reprocessing model, \citet{Cai2018} propose that the interaction between the local temperature fluctuation and a common large-scale fluctuation can naturally generate the AGN lags across UV/optical to X-ray. A scene where both disk and corona are coupled through magnetic fields provides an attractive physical mechanism responsible for the common large-scale fluctuation \citep{Sun2020ApJ...891..178S}.

This new scenario introduces a novel origin for the AGN inter-band lag as a result of the differential regression capability among distinct disk regions. When responding to a common large-scale fluctuation, the local fluctuation in the inner disk region generating emission at the shorter wavelength regresses quicker, due to the shorter local damping timescale of the temperature fluctuation, than that in the outer disk region radiating at the longer wavelength. Therefore, the implied AGN continuum at the longer wavelength naturally lags that at the shorter wavelength, nicely consistent with the lag measurements on local Seyfert galaxies \citep{Cai2018,Cai2020}. Besides, extending to X-ray, this framework also successfully accounts for the puzzling large UV to X-ray lags found in several local galaxies \citep{Cai2020}. 

As \citet{Cai2020} have pointed out, the random turbulence in the EUCLIA model could naturally yield randomness in lag measurements, and even reverse lags, i.e., UV leading X-ray instead of X-ray leading UV, as found in Mrk 509 \citep[][if dividing their LCs into two parts]{Edelson2017ApJ...840...41E} and Mrk 335 \citep[][comparing observations on the high state in 2008 and the low state in 2020, see their Figure~7]{Kara2023ApJ...947...62K}.

Comparing the thermal fluctuation model EUCLIA quantitatively to all individual AGNs with available lag measurements will be detailed in a companion paper (Z. B. Su et al., 2024a, in preparation).
In this work we make a prediction on the randomness of the optical inter-band lags for AGNs with similar BH mass and Eddington ratio to NGC 5548 (hereafter, N5548-like AGNs; see Section~\ref{sect:n5548_number} for details).
NGC 5548 is a Seyfert 1 galaxy at $z$ = 0.01067 with $M_{\rm BH} \simeq 5 \times 10^7~M_{\odot}$ and $\lambda_{\rm{Edd}} \simeq 0.05$ and has been largely used to calibrate our model. See \citet{Cai2018,Cai2020} for details and model parameters.

\subsection{Reprocessing model}\label{sect:reprocessing_model}

The conventional interpretation on the inter-band lags among UV/optical light curves is that the variability of UV/optical emission originates from the accretion disk heated by a varying ionizing source above the central BH, and the lags correspond to the time difference of the reverberation signals reflected between different disk regions \citep{Cackett2021}. 
To overcome the too-small lag predicted by the traditional reprocessing model, contamination from other more distant reprocessors, such as clouds in the BLR, has been proposed \citep[][]{Cackett2018, Chelouche2019, Montano2022, Netzer2022,Hagen2024}.
In the following, we adopt the simplest reprocessing model with DCE\footnote{Other reprocessing-related models, such as windy disk \citep{Sun2019MNRAS}, rimmed/tilted accretion disk \citep{Starkey2023}, and inward disk propagation with reverberation from the disk and wind \citep{Hagen2024}, are worthy of further investigation.} to simulate the optical LCs by convolving a driving LC with given response functions, which account for both the reverberation signal from disk and contributions of the DCE in different bands.

Suggested by \citet{Gardner2017} and following \citet{Zhu2018}, we assume that the driving LC is described by the damped random walk (DRW) model with parameters determined from modeling the observed {\it Swift}-UVW2 LC of NGC 5548. 
We adopted a damping timescale of 94.8 days and a mean flux of $2.527 \times 10^{-14}{\rm ~erg~cm^{-2}~s^{-1}~{\AA}^{-1}}$ determined by \citet{Zhu2018} using a $\sim$2-year UVW2 LC, while a variability amplitude of $0.419\times10^{-14}{\rm ~erg~cm^{-2}~s^{-1}~{\AA}^{-1}}$ (smaller by a factor of 1.77 than determined by \citealt{Zhu2018} but equivalent to the fractional variability of 0.166 determined by \citealt{Fau2016}) in order to match the observed $ugriz$ variability amplitudes of NGC 5548 measured for the same $\sim$6-month monitoring (see Section~\ref{sec:simulation_setting}).

Following \cite{Cackett2022ApJ} and \cite{Jaiswal2023}, we assumed $\psi_{\rm total}(\tau|\lambda) = (1-f) \psi_{\rm disk}(\tau|\lambda) + f \psi_{\rm BLR}(\tau)$ as the total response function accounting for the reverberation signals from both the disk\footnote{\url{https://github.com/drds1/astropy_stark/blob/master/astropy_stark/mytfb_quick.py}}, $\psi_{\rm disk}$ \citep[][see their Equations 19-23]{Starkey2017}, and clouds in BLR, $\psi_{\rm BLR}$ \citep[][see their Equation 5]{Cackett2022ApJ}, i.e., $\psi_{\rm BLR}(\tau|S, M)=1/S\tau\sqrt{2\pi}\times{\rm exp[-({\rm ln}~\tau-{\it M})^2/2{\it S}^2}]$, where $M \simeq {\rm ln}(7.3$/days) and $S = 1.1$ are the best-fit values for NGC 5548 from \citet{Cackett2022ApJ}.
Values for the response fraction from the BLR, $f$, also come from fits to the lag-frequency spectra of NGC 5548 by \citet[][their Figure~7]{Cackett2022ApJ}.

In this work, we adopt a simplified approach to model the effect of the BLR continuum emission on the lags. Our analysis has not yet considered other known BLR-related phenomena, such as the ``breathing'' effect \citep{Cackett2006} and the ``holiday'' anomaly \citep{Goad2016}, which can also contribute to the scatter in the lags to be discussed in the following and should be incorporated into a more physical BLR model in the future.

\subsection{Simulation setting}\label{sec:simulation_setting}

For the aforementioned two scenarios for AGN variability, we simulate both ideal and mock LCs in the five WFST-{\it ugriz} bands for 200 N5548-like AGNs. The ideal LCs span 10 yrs with a very fine sampling cadence of 0.1 day and without photometric uncertainty, while the mock LCs are obtained by sampling the ideal ones to which observational conditions are complemented as real as possible.

As a result of the season gap, some AGNs can only be continuously observed for a limited duration throughout the year (hereafter, the duration, $\mathcal{M}$, in units of month), though they can be repeatedly monitored for several years (hereafter, the baseline, $\mathcal{Y}$, in units of year). Furthermore, sparse (longer than $\sim$day) and irregular sampling cadences (hereafter, the cadence, $\mathcal{C}$, in units of day) are popular in the current time domain surveys for AGNs, whose photometry is affected by the measurement error (hereafter, the photometric uncertainty, $\sigma_{\rm e}$, in units of mag).

Thanks to the AGN STORM project \citep{Fau2016}, NGC 5548 was intensively monitored ($\mathcal{C} \simeq 1$) by ground-based telescopes in the SDSS-$ugriz$ bands over $\simeq 6$ month duration ($\mathcal{M} \simeq 6$) within one year ($\mathcal{Y} = 1$).
According to the $ugriz$ LCs of \citet{Fau2016}, the $u$-, $g$-, $r$-, $i$-, and $z$-band median apparent magnitudes are $\simeq 13.98$ mag, $\simeq 13.93$ mag, $\simeq 13.13$ mag, $\simeq 13.14$ mag, and $\simeq 12.85$ mag with median $\sigma_{\rm e}$ (including both measurement and calibration uncertainties in units of magnitude) of $\simeq 0.038$ mag, $\simeq 0.036$ mag, $\simeq 0.034$ mag, $\simeq 0.023$ mag, and $\simeq 0.013$ mag, respectively. 
Following \citet[][their Equation 8]{Vaughan2003MNRAS}, we find that for the duration of $\sim$6 months the observed variation amplitudes,  $\sigma_{\rm rms}$, after removing the contamination from host galaxy, for NGC 5548 are $\simeq 0.13$ mag, $\simeq 0.10$ mag, $\simeq 0.06$ mag, $\simeq 0.08$ mag, and $\simeq 0.05$ mag, for the $u$, $g$, $r$, $i$, and $z$ bands, respectively. 

For comparison, the means and standard deviations of our $\sigma_{\rm rms}$ simulated for the same duration of $\sim$6 months under the thermal fluctuation model are $0.11\pm 0.03$ mag, $0.10 \pm 0.03$ mag, $0.08 \pm 0.02$ mag, $0.07 \pm 0.02$ mag, and $0.06 \pm 0.02$ mag for the $u$, $g$, $r$, $i$, and $z$ bands, respectively. For the reprocessing model with DCE\footnote{The variation amplitude predicted by the reprocessing model is generally rescaled to match the observed value. Although the rescaling does not affect measuring the lag, the wavelength dependence of the variation amplitude is an important property of AGN variability and should be self-consistently addressed by future sophisticated reprocessing models.}, the resultant $\sim$6-month $\sigma_{\rm rms}$ is $0.10\pm 0.04$ mag, nearly the same for all $ugriz$ bands. Thus, for both scenarios, our simulations for N5548-like AGNs result in $\sigma_{\rm rms}$ comparable to those measured in $\sim$6 months for NGC 5548. 

The variation significance, SNR$_{\sigma} \equiv \sigma_{{\rm rms}} / \sigma_{{\rm e}}$, is a more important factor than $\sigma_{{\rm e}}$ in determining whether the inter-band lags can be successfully measured. We find that the observed SNR$_{\sigma}$ for NGC 5548 are $\simeq 3.3$, 2.6, 1.8, 3.0, and 3.4 for the $u$, $g$, $r$, $i$, and $z$ bands, respectively. Therefore, we would conservatively adopt SNR$_{\sigma} \simeq 3$ as a reference for all bands in our fiducial simulations and discuss how worse and better photometric uncertainties, i.e., SNR$_{\sigma} \simeq 1$ and 9, would affect the results.

To mimic the real observations, we assume that when sampling an ideal LC the observed epochs are randomly located within four hours before and after midnight. Besides, for a given ${\rm SNR}_\sigma$, magnitudes at all sampled epochs are fluctuated by random Gaussian deviates according to a specific photometric uncertainty of $\langle \sigma_{\rm rms} \rangle / {\rm SNR}_\sigma$, where $\langle \sigma_{\rm rms} \rangle$ is the mean of variation amplitudes of all simulated LCs for a given band.

\section{The inter-band lag of AGN}\label{sect:predit_lag}

\subsection{Does the inter-band lag vary?}\label{sec:ideal_lag_wavelength}

\begin{figure*}[ht!]
\centering

\includegraphics[width=0.45\textwidth]{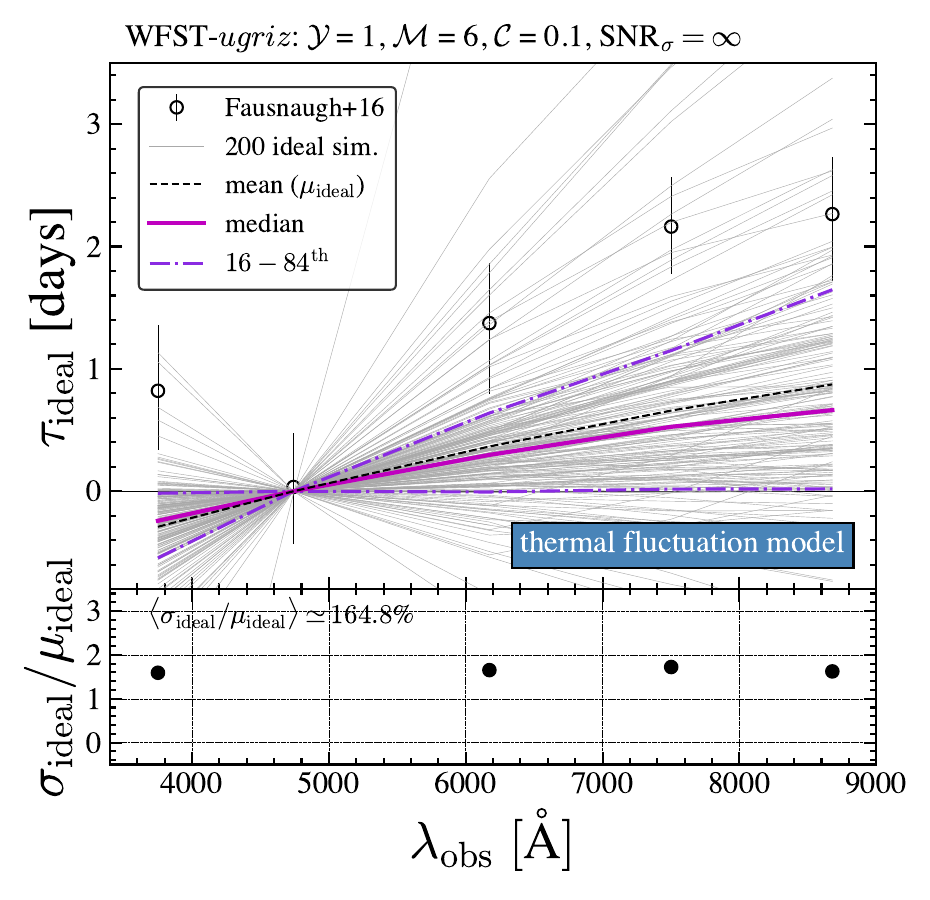}
\includegraphics[width=0.45\textwidth]{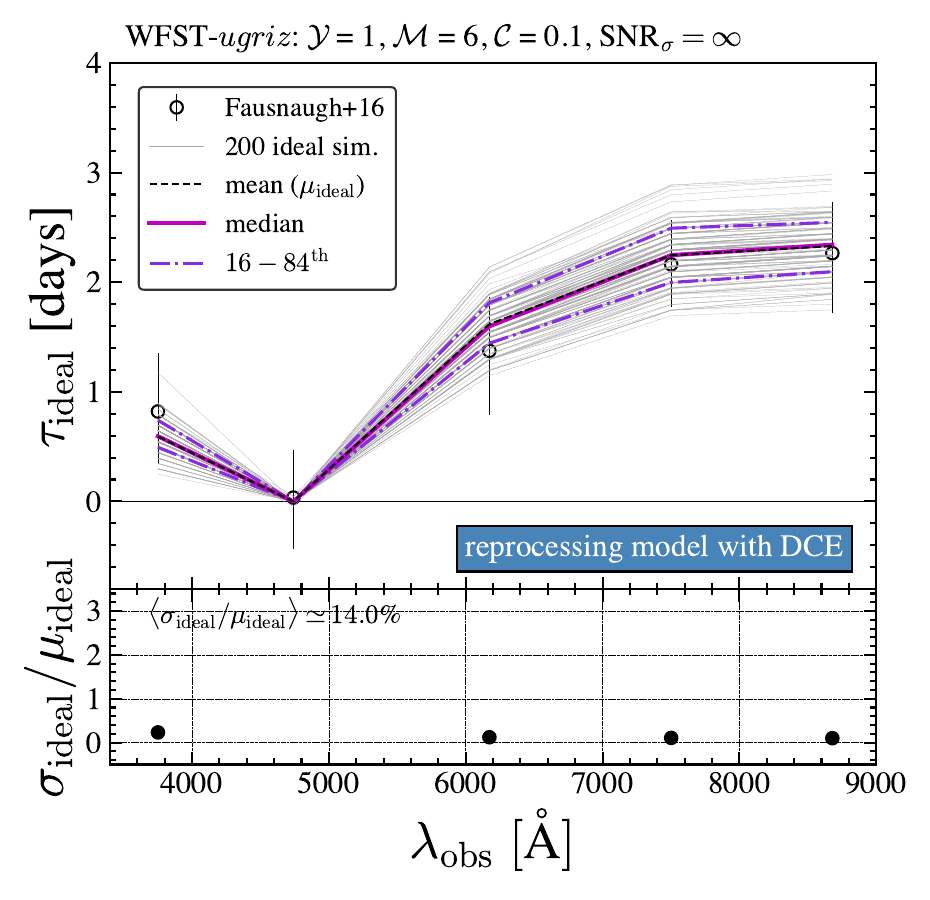}

\caption{
Left panel: relative to the WFST-$g$ band, the lag-wavelength relations for N5548-like AGNs predicted by the thermal fluctuation scenario are compared to the measured one (open circles) adopting the $\sim 6$ month $ugriz$ LCs of NGC 5548 from \cite{Fau2016}.
The predicted lag-wavelength relations are derived from 200 independent ideal simulations (gray thin solid lines) for N5548-like AGNs monitored over 6-month duration ($\mathcal{M}=6$) within one year ($\mathcal{Y}=1$) and with a very fine cadence of 0.1 day ($\mathcal{C}=0.1$) in the five WFST-$ugriz$ bands. Here considered is the ideal case without photometric uncertainty ($\sigma_{\rm e}=0$ or SNR$_\sigma = \infty$).
Accordingly shown are the median (thick solid line), mean (thin dashed line), and 16\%-84\% percentile ranges (thin dot-dashed lines) of the 200 ideal lag-wavelength relations. 
Although the 1$\sigma$ dispersion of lags, $\sigma_{\rm ideal}$, is larger at longer wavelength, similar relative scatters of lags are found for all bands as indicated by similar $\sigma_{\rm ideal}/\mu_{\rm ideal}$ (filled circles in the lower panel), where $\mu_{\rm ideal}$ is the corresponding mean lag.
Right panel: same as the left one, but for the prediction of the reprocessing model with DCE. We note that the relative scatters of the lags predicted by the reprocessing model with DCE are smaller than the thermal fluctuation model.}
\label{fig:agn_ideal_simulations}
\end{figure*}

Adopting $\mathcal{Y} = 1$, $\mathcal{M} = 6$, $\mathcal{C} = 0.1$, and SNR$_\sigma = \infty$ (i.e., $\sigma_{\rm e} = 0$), we first demonstrate that varied lags are predicted by both the thermal fluctuation model and the reprocessing model with DCE. The variation of the lags is intrinsically attributed to the random nature of AGN variability, although in reality both the sparse sampling and uncertainty of measurements would contribute the variation to some extent.
We adopt the physical parameters of NGC 5548, apply them to both AGN variability models, and perform 200 independent simulations to mimic WFST observations on 200 N5548-like AGNs.
For the simulated WFST-$ugriz$ LCs of every AGN, we use the standard interpolated cross-correlation function (CCF) technique\footnote{PyCCF: https://www.ascl.net/1805.032} (e.g., \citealt{Peterson1998, PyCCF2018}) to measure the inter-band lags relative to the $g$ band. We consider a broad enough range spanning -50 days to 50 days to search the CCF for the centroid lag, $\tau$, defined by CCF-weighted averaging the lags whose correlation coefficients, $r_{\rm{cc}}(\tau)$, are larger than 80\% of the maximum. Then, there is a lag-wavelength relation for every N5548-like AGN.
Note that, besides the CCF method adopted here, other methods, such as the frequency-resolved technique \citep{Zoghbi2013, Cackett2022ApJ} and the wavelet transform method \citep{Wilkins2023}, may be helpful in unveiling the nature of the AGN lags.

The lag-wavelength relations of 200 N5548-like AGNs predicted by the thermal fluctuation model and the reprocessing model with DCE are shown as gray solid lines in the left and right panels of Figure~\ref{fig:agn_ideal_simulations}, respectively. 
The predicted lag-wavelength relations are intriguingly diverse for both models. 
It is easy to notice that both the lags and the lag scatters are statistically larger at longer wavelength, but the relative scatter of the lags, defined as the ratio of the lag scatter to the mean lag, are nearly identical across wavelengths. Given $\mathcal{M} = 6$ and $\mathcal{Y} = 1$, the thermal fluctuation model implies a relative scatter in the lags of $\sim 165\%$, which is significantly larger than $\sim14\%$ implied by the reprocessing model with DCE. 
For comparison, the reprocessing model without DCE (i.e., disk only) implies a relative scatter in the lags of only $\sim7\%$.
Note that the relative scatter of the lags decreases with increasing the duration/baseline of LC.

For the reprocessing models, the CCF between two bands is related to the auto-correlation function (ACF) of the driving LC and to the given transfer functions of the two bands.
Since the ACF of the driving LC is determined by the power spectral density \citep[e.g.,][]{Edelson1988, Welsh1999}, which changes from one finite duration to another owing to the random nature of AGN variability, the resultant is a scatter in the lag-wavelength relations. 
Furthermore, the reprocessing model with DCE, involving more distant reverberation signals from BLR and thus longer light-traveling times, results in a larger scatter in the lag-wavelength relations than the reprocessing model without DCE. 

In contrast, the inter-band lags in the thermal fluctuation scenario are attributed to the differential regression capability of local fluctuations, which is likely related to the thermal timescales. 
Thus the complicated origin of the lags in the thermal fluctuation model is expected to result in a larger scatter in the lag-wavelength relations than the reprocessing model with DCE.

\begin{figure*}[!ht]
    \centering
    \includegraphics[width=0.9\textwidth]{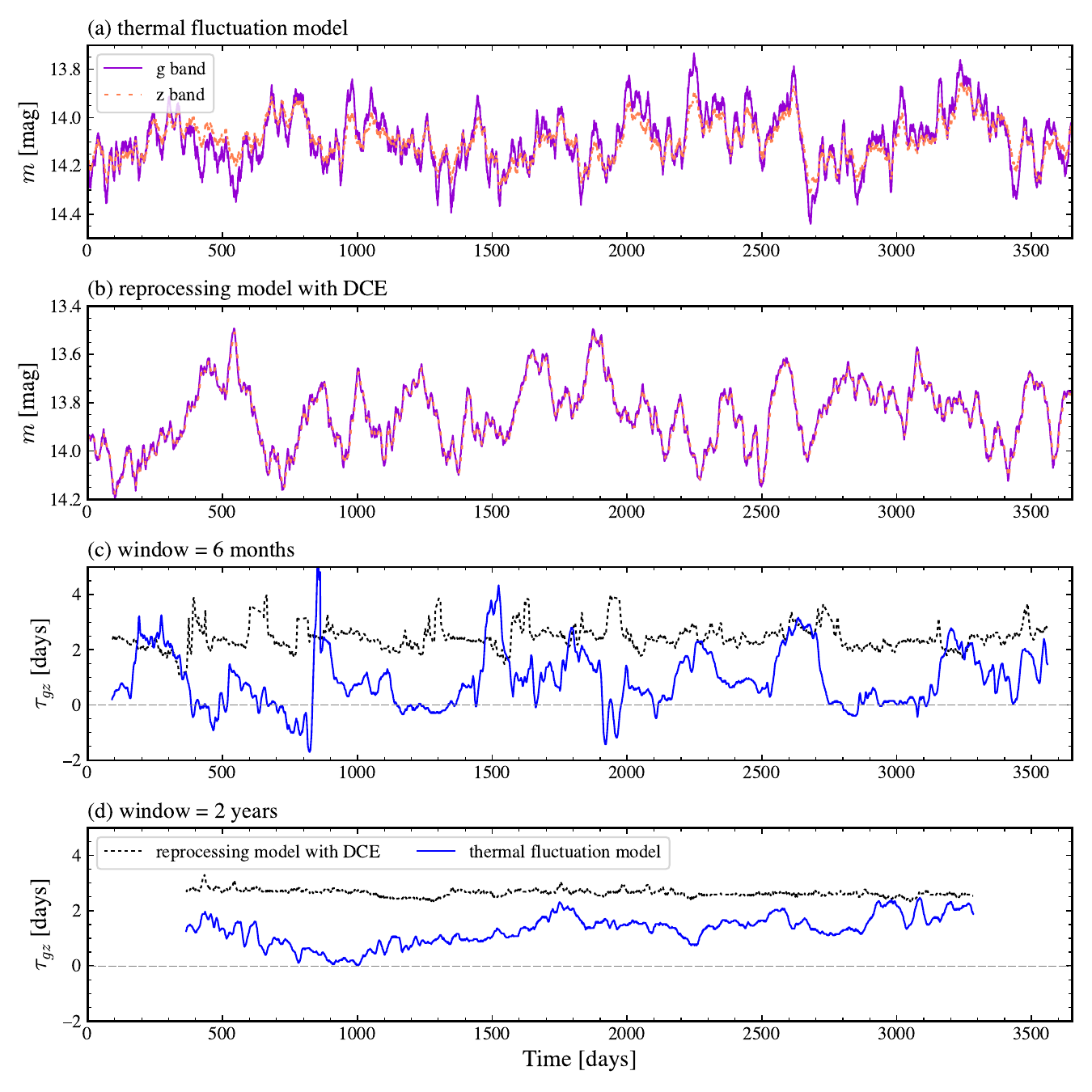}
    \caption{
    An illustration of the simulated 10-year ideal LCs ($g$ band versus $z$ band) for the thermal fluctuation model (panel a) and the reprocessing model with DCE (panel b). 
    Using part of LCs blanketed by a moving window of 6 months (or 2 years) to measure the inter-band lags between $g$ and $z$ bands, the evolution of $\tau_{gz}$ with time for both models are presented in panel c (or panel d). 
    Being consistent with the results shown in Figure~\ref{fig:agn_ideal_simulations}, the lag evolution implied by the thermal fluctuation model exhibits larger variation than the reprocessing model with DCE. With increasing the size of the moving window, the lag evolution becomes less variable for both models.}
    \label{fig:agn_lag_with_time}
\end{figure*}

The potential change of the lags observed in a few AGNs (see Introduction and Figure~\ref{fig:NGC4395_lag_test}) may support our simulation results, but more observational tests are demanded. 
Using long-term photometric surveys like ZTF, there are hints at the lag change for individual AGNs (Z. B. Su et al., 2024b, in preparation).
It is interesting to note that a large scatter in the ratios of the measured lags to the SSD-predicted lags was reported in a sample of 95 quasars from SDSS-RM project \citep[e.g.,][]{Homayouni2019, Sharp2024}, which may also hint at the lag randomness.

\subsection{How do the inter-band lags evolve with time?}

In Figure~\ref{fig:agn_lag_with_time}, we present distinguishable predictions of both the thermal fluctuation model and reprocessing model with DCE on the lags evolution. 
We find that the correlation time for a significant lag change differs from the damping timescale (or auto-correlation timescale) of a single-band LC, and depends on the duration of the LC used to measure the lags. 
A longer duration of the LC results in a smoother lag evolution, indicating a longer correlation time for the lag change.
Thus, how the lags evolve with time and what is the relationship between the lag evolution behavior and the chosen duration/baseline of the LC are new questions to be addressed by future monitoring of AGNs.

\subsection{Is there always a $u$-band excess?}

In the 6-month campaign for NGC 5548 by \citet{Fau2016}, there is a clear $u$-band excess in the lag-wavelength relation, that is, the $u$-band variation delays rather than leads the $g$-band variation (Figure~\ref{fig:agn_ideal_simulations}). 
The $u$-band excess has been observed in many AGNs and is widely thought to be the result of contamination from the diffuse BLR emission \citep[e.g.,][]{Cackett2018, Lawther2018, Korista2019, Chelouche2019, Netzer2022, Montano2022, GuoH2022}. 
However, for several AGNs, repeated observations on Mrk 110 show that the $u$-band excess does not always exist \citep{Vincentelli2022}. 
For a sample of 22 quasars, \citet{Yu2020} claimed minimal contamination from the diffuse BLR emission on the lags, and \citet{Sharp2024} suggested a lack of evidence for contributions of the diffuse BLR emission to the lags in 95 luminous quasars.

Here in Figure~\ref{fig:agn_ideal_simulations}, we show that the $u$-band excess is a distinguishable prediction for the thermal fluctuation model and reprocessing model with DCE.
In the left panel of Figure~\ref{fig:agn_ideal_simulations}, the median/mean lag-wavelength relation implied by the thermal fluctuation model is monotonic, although the $u$-band excess is found in a few simulations.
Instead, the $u$-band excess always exists in each simulation based on the reprocessing model with DCE (the right panel of Figure~\ref{fig:agn_ideal_simulations}).
Expectedly, future long-term photometric surveys on AGNs, especially quasars with little diffuse BLR emission suggested by \citet{Yu2020} and \citet{Sharp2024}, would shed new light on the nature of the $u$-band excess.

\subsection{Are there correlations between the lags and variation properties?}\label{sect:lag_variation_properties}

\begin{figure*}[t!]
    \centering
    \includegraphics[width=0.8\textwidth]{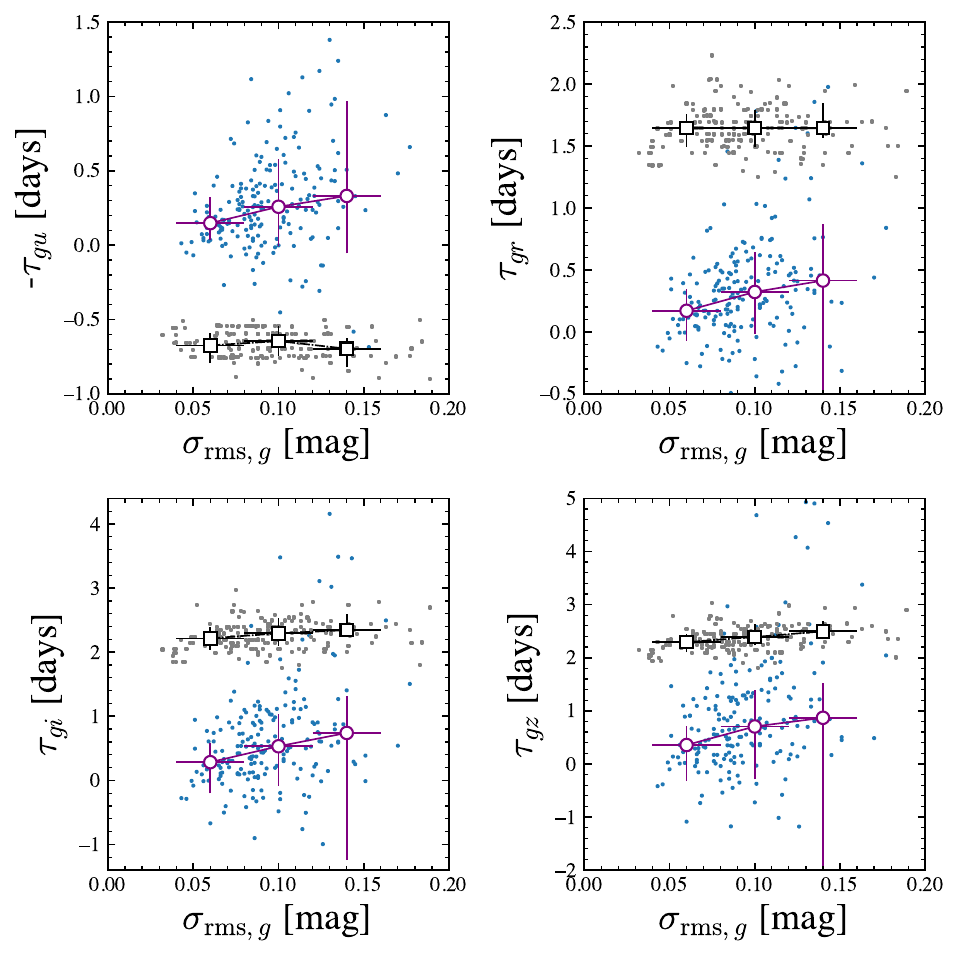}
    \caption{Correlations between the lag of the $x$ band relative to the $g$ band, $\tau_{gx}$, and the $g$-band variation amplitude, $\sigma_{{\rm rms,}g}$, where $x$ takes $u$ (top-left panel), $r$ (top-right panel), $i$ (bottom-left panel), and $z$ (bottom-right panel), respectively. 
    Values of $\tau_{gx}$ and $\sigma_{{\rm rms,}g}$ from individual simulations are depicted by the small solid symbols, while the corresponding median lag and 16\%-84\% percentile ranges are shown as the large open symbols with vertical bars in even bins of 0.04 mag. Only bins containing more than ten data points are shown. Circles are for the thermal fluctuation model, while squares are for the reprocessing model with DCE.
    The thermal fluctuation model predicts a clear positive correlation between the variation amplitude and the lag (i.e., the larger the variation amplitude, the longer the lag), while the reprocessing model with DCE does not predict such a correlation.
    }
    \label{fig:agn_var_lag}
\end{figure*}

A potential positive correlation between the lag and variation amplitude has been observed in NGC 4395 \citep[][]{Montano2022,McHardy2023MNRAS.519.3366M}. To examine this correlation, we display in Figure \ref{fig:agn_var_lag} the relationship between the lag and variation amplitude implied by both the thermal fluctuation model and reprocessing model with DCE. Interestingly, a positive correlation is predicted by the thermal fluctuation scenario, while the reprocessing model with DCE does not predict such a correlation. 

For the reprocessing model with DCE, we find that assuming different variation amplitudes for the driving LC results in more or less the same values of the mean/median lag and scatter in the lags. This explains why the reprocessing model with DCE does not predict a correlation between the inter-band lag and the variation amplitude. 
Instead, we find that for the reprocessing model with DCE the lags highly depend on the damping timescale of the driving LC, that is, a shorter damping timescale generally leads to a smaller inter-band lag. 
A similar result has been reported by investigating the lags between the broad emission line and the ionizing UV continuum \citep[][see their Figure~9]{Goad2014}. Therefore, future measurements on the correlations between the lags and variation properties (e.g., timescale and amplitude) would be useful in distinguishing models for AGN variability.

\subsection{How to infer model parameters from the varied lags?}\label{sec:mock_lag_wavelength}

If the lags do indeed vary, averaging the lags measured in repeated and non-overlapping baselines would be necessary before inferring the model parameters.
Adopting the thermal fluctuation scenario, we present how the sparse sampling ($\mathcal{C} = 1$) and the moderate photometric uncertainty (SNR$_\sigma \simeq 3$) would add scatter to the intrinsic dispersion of the lags.
Then, we discuss to what extent the lag-wavelength relation can be accurately measured by considering cases of various sampling cadences ($\mathcal{C} = 3$ or 5) and photometric uncertainties (SNR$_\sigma \simeq 1$ or 9).

\begin{figure}[t!]
\centering
\includegraphics[width=0.45\textwidth]{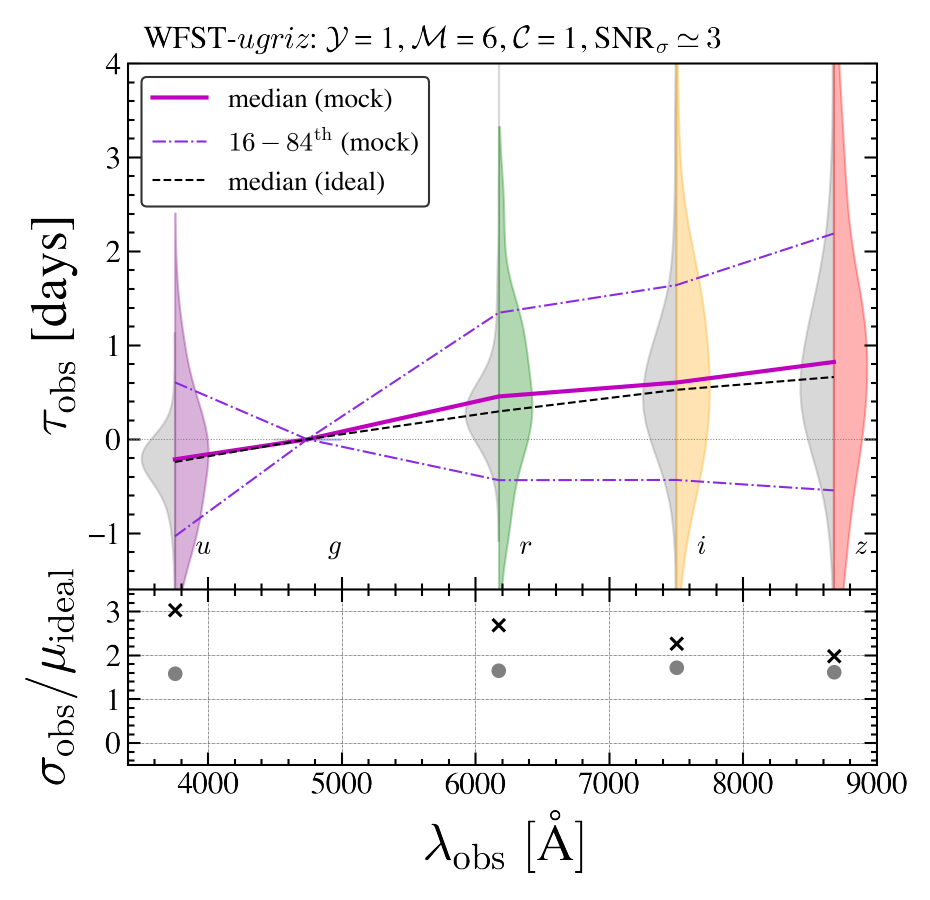}
\caption{
Illustration on how sparse sampling and photometric uncertainty would add scatter to the intrinsic dispersion of the lag in terms of the thermal fluctuation model. In each band, the left portion of the violin plot displays the distribution of 200 ideal lags inferred from the ideal LCs with a very fine cadence and without photometric uncertainty (Figure~\ref{fig:agn_ideal_simulations}), while the right portion shows that of 200 mock lags inferred from the mock LCs with sparse sampling ($\mathcal{C} = 1$) and moderate photometric uncertainty (SNR$_\sigma \simeq 3$). 
The median (thick solid line) and 16\%-84\% percentile ranges (thin dot-dashed lines) of the mocked lags are compared to the median (thin dashed line) of the ideal lags.
This comparison suggests that the median lag-wavelength relations inferred from the mock and ideal LCs are comparable, regardless of the sparse sampling and photometric uncertainty, both of which indeed increase the dispersion of the lag. Dispersion of the mock lags (crosses) and ideal lags (circles) are compared in the lower panel.
}

\label{fig:agns_observed}
\end{figure}

\begin{figure}[t!]
    \centering
    \includegraphics[width=0.45\textwidth]{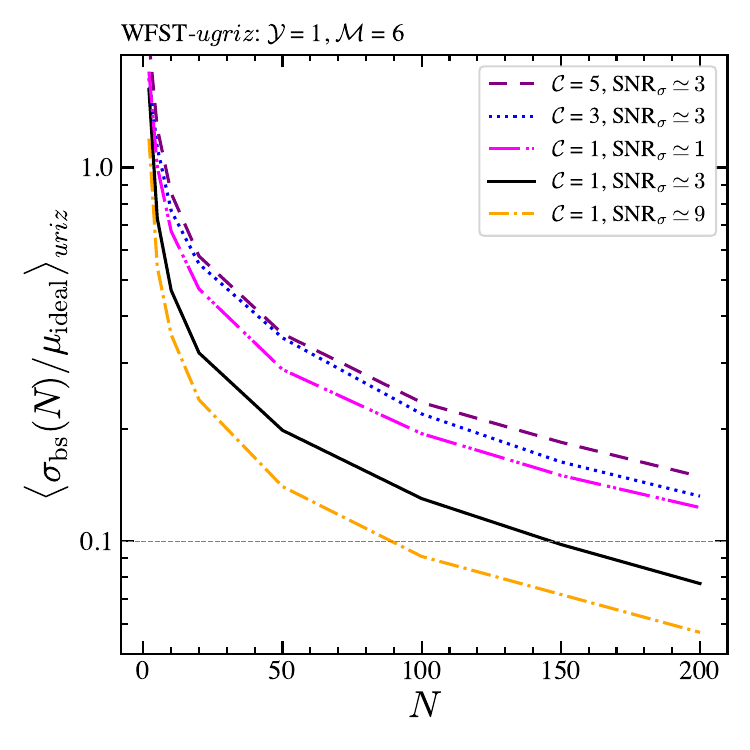}
    \caption{For the thermal fluctuation model,
    the accuracy of the true lag-wavelength relation obtained by averaging the observed lag-wavelength relations of $N$ AGNs with comparable physical properties for different sampling cadences ($\mathcal{C}$ $\simeq$ 1, 3, 5) and photometric uncertainties (SNR$_\sigma \simeq 1,3,9$).}
    \label{fig:agn_average}
\end{figure}

In Figure~\ref{fig:agns_observed}, the colored right portion and gray left portion of the violin plot represent distributions of 200 mock and ideal lags inferred from the mock and ideal LCs, respectively.
By comparing dispersions of the ideal lags (the left panel of Figure~\ref{fig:agn_ideal_simulations} or $\sigma_{\rm{ideal}}/\mu_{\rm{ideal}}$) and the mock lags (Figure~\ref{fig:agns_observed} or $\sigma_{\rm{obs}}/\mu_{\rm{ideal}}$), the dispersion of the lags is larger as expected once considering sparse sampling and photometric uncertainty. 
However, the median values of the mock and ideal lags are comparable (Figure~\ref{fig:agns_observed}). This implies that averaging the lag-wavelength relations of hundreds of AGNs with comparable physical properties could be an efficient way of obtaining the true lag-wavelength relation.

To assess to what extent the true lag-wavelength relation can be accurately measured after averaging the lag-wavelength relations of some AGNs with comparable physical properties, we illustrate in Figure~\ref{fig:agn_average} the accuracy quantified by $\langle\sigma_{\rm{bs}}(N)/\mu_{\rm{ideal}}\rangle_{uriz}$ as a function of the used AGN number, $N$. 
On the basis of the lag-wavelength relations of 200 N5548-like AGNs, we average $N$ of them (randomly selected with replacements) for a mean lag-wavelength relation and repeat the selection plus average for 1000 mean lag-wavelength relations whose 1$\sigma$ dispersion is nominated as $\sigma_{\rm bs}(N)$. Finally, averaging the ratios of $\sigma_{\rm bs}(N)$ to $\mu_{\rm{ideal}}$ over $uriz$ bands, i.e., $\langle\sigma_{\rm{bs}}(N)/\mu_{\rm{ideal}}\rangle_{uriz}$, represents the global accuracy achievable by averaging the lag-wavelength relations of $N$ AGNs.
Averaging the lag-wavelength relations of $N \sim 20$ AGNs only, an accuracy as high as $\sim 50\%$ can be easily achieved, but the accuracy increases mildly with further increasing the AGN number, e.g., reaching $\sim 20\%$ for $N \sim 100$. Instead, increasing finer sampling cadence ($\mathcal{C}$ $\simeq$ 1) and/or decreasing photometric uncertainty (SNR$_\sigma \simeq 9$) are more efficient in obtaining high accuracy for $\sim 100$ AGNs.
On the other hand, high accuracy could be achieved even under worse conditions, such as a long sampling cadence ($\mathcal{C}$ $\simeq$ 5) and/or low significance of variation (SNR$_\sigma \simeq 1$), as long as a large number of similar AGNs ($N > 200$) are available.

\section{Forecasts for the WFST survey}\label{WFST}

To serve as an indispensable complement on the northern sky to the southern LSST surveys, the WFST, located on the summit of Saishiteng Mountain near Lenghu in northwestern China, started the engineering commissioning observations in mid-August 2023. 
The Deep High-cadence $u$-band Survey (DHS), one of the key programs planned for the WFST 6-year survey \citep{wfst2023}, would cover $\sim$ 720 deg$^2$ surrounding the equator and blanketing part of the SDSS Stripe 82 region. Two separate DHS fields of $\sim 360$ deg$^2$ will be continuously monitored for 6 months per year in $ugri$ bands and daily cadence (except $u$ around the full moon and $i$ around the new moon).
Besides, both the whole COSMOS field and an area of $\sim 10$ deg$^2$ surrounding the North Ecliptic Pole (NEP) would be monitored in $ugri$ bands every night and last for 6 months per year and $\gtrsim 9$ months per year, respectively.
Therefore, these WFST surveys are valuable for studying the inter-band lags of AGNs and examining our prediction on lag randomness.
To suggest an optimal strategy for the inter-band lag of AGNs with the upcoming formal WFST survey, we perform a series of simulations to assess the effects of diverse observational conditions on the lag measurement based on the thermal fluctuation model, including band selection (e.g., $ugri$ or $gri$), sampling cadence (e.g., 2 visits per night or 1 visit per several nights), the variation significance (SNR$_\sigma \simeq 1$, 3, or 9), duration ($\mathcal{M} = 3$, 6, or 9), and baseline ($\mathcal{Y}=1,$ 2, or 6).

In this section, to quantify the global performance on retrieving the inter-band lags involving several (at least two) bands and compare results implied by different observational conditions, we fit the commonly used function form, i.e., $\tau = \tau_g [(\lambda/\lambda_g)^\beta - 1]$ with a fixed $\beta$ of 4/3 predicted by the SSD model \citep[e.g.,][]{Fau2016,Mudd2018}, to our mock lag-wavelength relations, which are measured relative to the $g$ band with an effective wavelength of $\lambda_g$, and derive the $g$-band lags, $\tau_g$, which are assumed to be relative to the X-ray emission of corona located at $\tau = 0$. A positive (negative) $\tau_g$ indicates that the X-ray variation leads (lags) the $g$-band variation.
On one hand, using the CRM method to estimate the disk size, the absolute size of the disk at $\lambda_g$ is simply taken to be $c \tau_g$ where $c$ is the speed of light. For a negative $\tau_g$, it would result in a weird negative disk size and is generally discarded. For example, \cite{GuoH2022} got rid of negative lags ($< 20\%$ in their initial sample) inferred from the CCF analysis between the ZTF $g$- and $r$-band LCs. However, negative lags are indeed predicted by the thermal fluctuation scenario (see Figure~\ref{fig:agn_ideal_simulations} for ideal simulations). Thus we decide to keep the negative lags inferred from our mock simulations and directly use $\tau_g$ for statistics rather than the disk size, $c \tau_g$.
On the other hand, as we discussed in Section~\ref{sec:accretion_disk}, the foundation of the CRM method may be questionable. Here the derived $\tau_g$ is solely adopted as a common proxy to quantify the effects of diverse observational conditions.

\begin{figure*}[t!]
    \centering
    \includegraphics[width=0.32\textwidth]{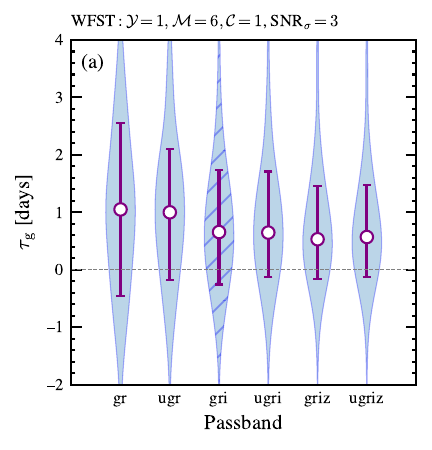}
    \includegraphics[width=0.32\textwidth]{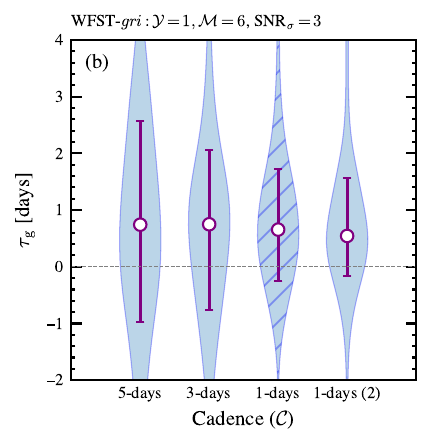}
    \includegraphics[width=0.32\textwidth]{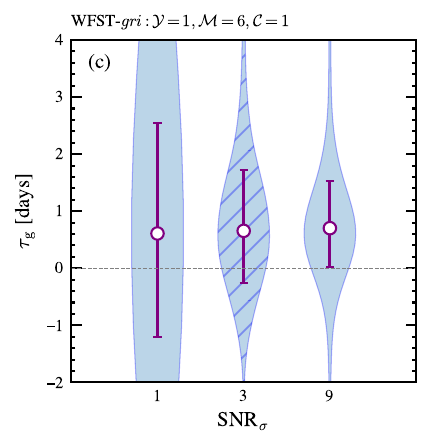}
    \includegraphics[width=0.32\textwidth]{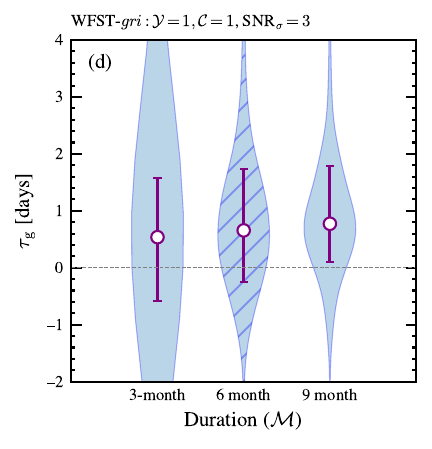}
    \includegraphics[width=0.32\textwidth]{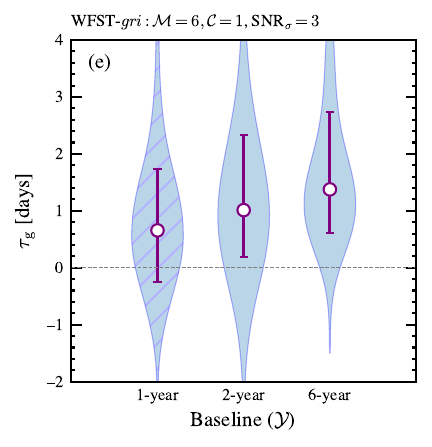}
    \caption{Effects of diverse observational conditions on the lag measurement in terms of the thermal fluctuation model. For each case, the $\tau_g$ distribution is shown as a violin plot, overlaid by the median (open circle) and 16\%-84\% percentile ranges (the vertical bar). The hatched violin plot in each panel refers to the reference case for a combination of \{WFST-$gri$, $\mathcal{C} = 1$, SNR$_\sigma = 3$, $\mathcal{M}=6$, $\mathcal{Y}=1$\}. Each panel illustrates the effect of solely changing one of the conditions compared to the reference case: (a) passband, (b) cadence, (c) variation significance, (d) duration, and (e) baseline.}
    \label{fig:strategy_all}
\end{figure*}

\subsection{Observational effects on the lag measurement}

\subsubsection{Passbands}\label{sec:passband}

For analyzing the inter-band lag, it is undeniable that conducting quasi-simultaneous observations in more passbands is a better option, but it requires a large amount of observational times. Panel (a) of Figure~\ref{fig:strategy_all} illustrates the impact of considering different combinations of WFST bands, i.e., $gr$, $ugr$, $gri$, $ugri$, $griz$ and $ugriz$, on the accuracy of measuring $\tau_g$. As the number of bands increases, the dispersion of measured $\tau_g$ decreases. 
However, the decreasing rate of the dispersion becomes quite slow once there are three or more bands involving $gr$. Furthermore, due to the larger time lag $\tau_{gi}$ compared to $\tau_{gu}$, the combination of $gri$ is expected to have a smaller dispersion compared to $ugr$. We note there are slightly systematic offsets in cases of using $gr$ and $ugr$ bands, implying that a combination of too few passbands (e.g., $gr$) or too narrow wavelength coverage (e.g., $ugr$) could result in a larger $\tau_g$. 
Therefore, involving the longer wavelength, e.g., the $i$ band, is important in constraining $\tau_g$. In short, we demonstrate that quasi-simultaneous observations in the $gri$ bands are preferred. Instead, the combination of $ugri$ is also appealing and involving the $u$ band can also reveal whether the diffuse continuum emission from the BLR can contaminate that of the accretion disk \citep[e.g,][]{Cackett2018, Montano2022, Netzer2022} or not \citep[e.g.,][]{Vincentelli2022, Sharp2024}.

\subsubsection{Sampling Cadence and Variation Significance}\label{sec:cadence_SNR}

In panel (b) of Figure~\ref{fig:strategy_all}, a cadence of approximately 3-5 days leads to a large dispersion in $\tau_g$, while a cadence more frequent than 1 day (i.e., 2 visits per night) does not significantly reduce the dispersion but instead requires double observational times. Therefore, a cadence of 1 day is optimal for the local N5548-like AGNs as well as high-redshift AGNs thanks to the cosmic time dilation.

In panel (c) of Figure \ref{fig:strategy_all}, it is clear that increasing SNR$_\sigma$ from 1 to 9 indeed reduces the dispersion in $\tau_g$. However, the dispersion does not reduce much when SNR$_\sigma$ increases from 3 to 9. This is because when the photometric uncertainty is very small, the dispersion in $\tau_g$ is intrinsic and dominated by the randomness of the lags, predicted by the thermal fluctuation scenario (Section~\ref{sect:predit_lag}).
Interestingly, even for SNR$_\sigma = 1$, the corresponding median lag does not differ much from that implied by SNR$_\sigma = 9$. This means that even for AGNs with smaller variation amplitude or larger photometric uncertainty, we may still unveil the true inter-band lag once averaging a large sample of AGNs with similar physical properties.

\begin{figure*}[t!]
    \centering
    \includegraphics[width=0.45\textwidth]{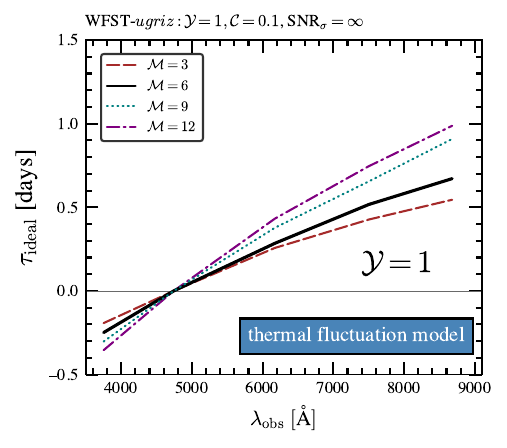}
    \includegraphics[width=0.45\textwidth]{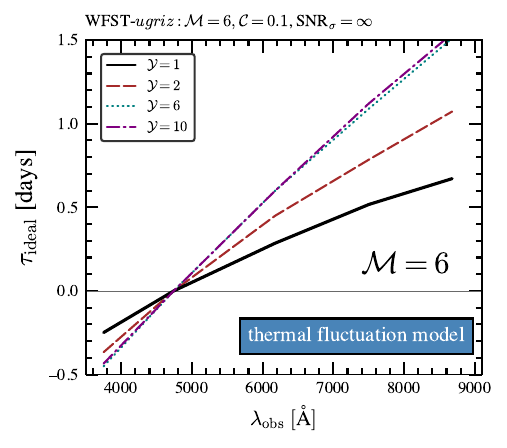}
    \includegraphics[width=0.45\textwidth]{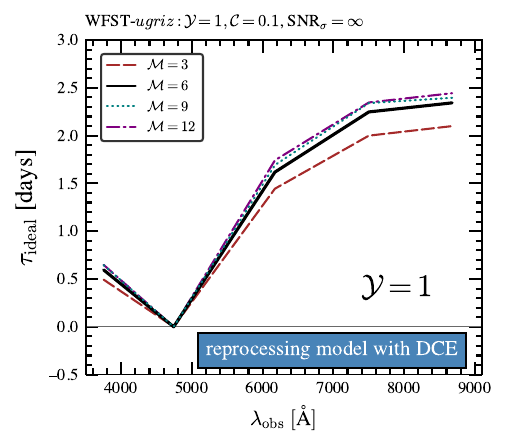}
    \includegraphics[width=0.45\textwidth]{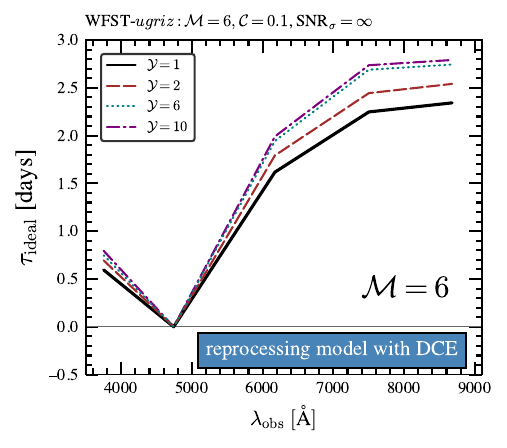}
    \caption{Top panels: taking the median of 200 ideal lag-wavelength relations inferred from the mock WFST-$ugriz$ LCs generated by the thermal fluctuation scenario, the median lag-wavelength relation steepens with increasing duration ($\mathcal{M}$ from 3 months, through 6 and 9, to 12 months in one year; top-left panel) and baseline ($\mathcal{Y}$ from 1 year, through 2 and 6, to 10 years with a fixed 6-month duration per year; top-right panel). Bottom panels: same as the top panels, but for the prediction of the reprocessing model with DCE.}
    \label{fig:agn_year}
\end{figure*}

\subsubsection{Duration and Baseline}\label{sec:duration_baseline}

In panels (d) and (e) of Figure~\ref{fig:strategy_all}, we show the impacts of duration and baseline on the lag measurements, respectively. 
A slight increase in $\tau_g$ with increasing either the duration (within one year) or the baseline (with a fixed duration of 6 months per year) is observed. 

To deeply explore the dependence of the lag-wavelength relation on both duration ($\mathcal{M} = 3$, 6, 9, and 12) and baseline ($\mathcal{Y} = 1$, 2, 6, and 10), we illustrate in Figure~\ref{fig:agn_year} the median lag-wavelength relations of 200 ideal lag-wavelength relations implied by both the thermal fluctuation model (top panels) and the reprocessing model with DCE (bottom panels).
Both models predict a steepening effect for the lag-wavelength relation with increasing duration and baseline. The lag-wavelength relation eventually gets saturation for a sufficiently long baseline. 
For the reprocessing model with DCE, the saturated lags are found to be consistent with the lags implied by weighting the transfer function adopted. Note that a similar result has been reported by investigating the lags between the broad emission line and the continuum \citep[][see Figure~8 therein]{Welsh1999}.

In fact, the measured variation timescale of a LC is biasedly small for a short duration or baseline. Assuming the DRW process, it has been suggested that the LC length should be at least 10 times the damping timescale such that the unbiased variation timescale can be retrieved \citep[e.g.,][]{Kozlowski2021AcA....71..103K, Hu2024}. In the reprocessing model with DCE, we adopt $\sim$95 days for the input damping timescale, thus it would require a LC length of at least $\sim$3 years to retrieve the unbiased variation timescale. Interestingly, according to the bottom panels of Figure~\ref{fig:agn_year}, the inter-band lags get saturation for a sufficiently long baseline (i.e., several years), comparable to that required to retrieve the unbiased variation timescale.
With the help of upcoming high-cadence and long-term WFST/LSST surveys, investigating the steepening effect of the lag-wavelength relation (Figure~\ref{fig:agn_year}) as well as the relationship between the inter-band lag and variation timescale (see also Section~\ref{sect:lag_variation_properties}) would shed new light on the underlying physics of AGN variability.

\begin{figure*}[t!]
\centering
\includegraphics[width=0.9\textwidth]{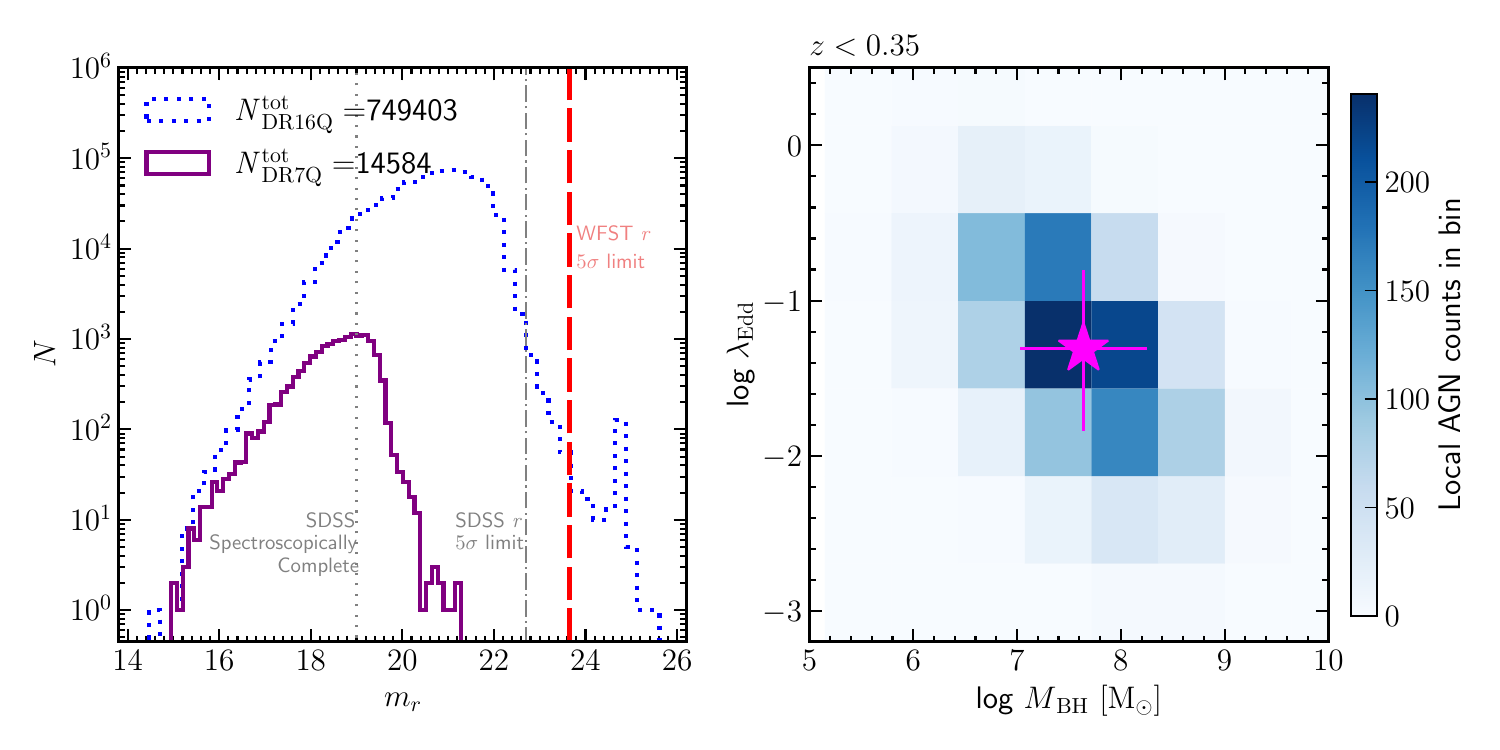}
\caption{Left panel: distributions of the apparent $r$-band magnitudes for spectroscopically confirmed quasars in the SDSS  DR16Q catalog (dotted histogram; \citealt{sdssdr16q}) and a uniform, complete sample of broad-line AGNs at $z < 0.35$ selected from the SDSS DR7Q catalog (solid histogram; \citealt{Liu2019}). Note that only quasars with physical $r$-band magnitudes are used here. Shown for comparison are the spectroscopically complete limit of $\sim$ 19 mag for the SDSS quasars (vertical dotted line) and the WFST $r$-band 5$\sigma$ detection limit of $\sim$ 23.7 mag in a single 90~s exposure \citep[vertical dashed line;][]{Lei2023}, which is deeper than that of the SDSS-$r$ band (vertical dot-dashed line).
Right panel: the density map in $\log M_{\rm{BH}}$ versus $\log \lambda_{\rm{Edd}}$ for local N5548-like AGNs expected in $\sim$ 1000 deg$^2$, to be monitored by the WFST DHS, by scaling that of the \citet{Liu2019} sample covering 9376 deg$^2$.
The star and horizontal (or vertical) bar indicate the median and 16\%-84\% percentile ranges of $\log M_{\rm{BH}}$ (or $\log \lambda_{\rm{Edd}}$) of the \citet{Liu2019} sample.}
\label{fig:agn_detection}
\end{figure*}

\subsection{The number of available local N5548-like AGNs}\label{sect:n5548_number}

In this work, we focus on the N5548-like AGNs and suggest that by taking ensemble average over many such AGNs one can derive a reliable underlying lag-wavelength relation. To reach a precision of $\sim 10\%$ for the lag-wavelength relation, we would need a few hundred N5548-like AGNs (Figure~\ref{fig:agn_average}). Here we show that these N5548-like AGNs are typical in the local Universe ($z < 0.35$) and there are a few hundred to be covered by the WFST surveys.
 
The SDSS survey, mainly covering the northern sky, has spectroscopically confirmed $\gtrsim 750,000$ quasars according to its Sixteenth Data Release quasar (DR16Q) catalog \citep[]{sdssdr16q}. Figure~\ref{fig:agn_detection} presents the distribution of the $r$-band magnitude for the DR16Q quasars (dotted histogram) overlapped by the SDSS spectroscopic complete limit (vertical dotted line) as well as the SDSS $r$-band 5$\sigma$ detection limit (vertical dot-dashed line). 
In a 90~s exposure, the WFST DHS is expected to reach a 5$\sigma$ detection limit in the $r$ band \citep[$\sim 23.7$ mag, vertical dashed line;][]{Lei2023}\footnote{The WFST $r$-band 5$\sigma$ detection limit in a 90~s exposure is calculated for conditions of airmass = 1.20, seeing = 0.75 arcsec at the darkest New Moon night of the Lenghu site ($V = 22.30$ mag, Moon phase $\theta=0^\circ$). The code for the detection limit is available at \url{https://github.com/Leilei-astro/WFST-limiting-magnitudes}.} deeper than that of the SDSS.
Although the WFST surveys would provide valuable long-term variation data (weekly cadence in the WFST wide-field survey) for a majority of the SDSS DR16Q quasars, only a portion of them located within the WFST DHS and at relatively low redshift are suitable for analyzing the inter-band lag. 

\citet{Liu2019} constructed a uniform, complete sample of broad-line AGNs in the local Universe ($z < 0.35$) selected from the SDSS Seventh Data Release quasar (DR7Q) catalog. The corresponding distribution of the $r$-band magnitude is shown as the solid histogram in the left panel of Figure~\ref{fig:agn_detection}. Additionally, the right panel of Figure~\ref{fig:agn_detection} shows the density map in $\log \lambda_{\rm{Edd}}$ versus $\log M_{\rm{BH}}$ by scaling the density map of the \citealt{Liu2019} sample in 9376 deg$^2$ to $\sim 1000$ deg$^2$. These local AGNs have typical $\log (M_{\rm{BH}}/M_\odot) \simeq 7.6$ and $\log \lambda_{\rm{Edd}} \simeq -1.3$ (i.e., the star in the right panel of Figure~\ref{fig:agn_detection}), indeed similar to NGC 5548. 
The number of available N5548-like AGNs to be covered by the WFST DHS is several hundred (and more if selecting similar AGNs from the SDSS DR16Q). Therefore, we expect that there are enough N5548-like AGNs in the WFST DHS valuable for analyzing the inter-band lag of AGN.

\subsection{WFST versus other time domain surveys}\label{projects}

\begin{figure}[t!]
\centering
 \includegraphics[width=0.45\textwidth]{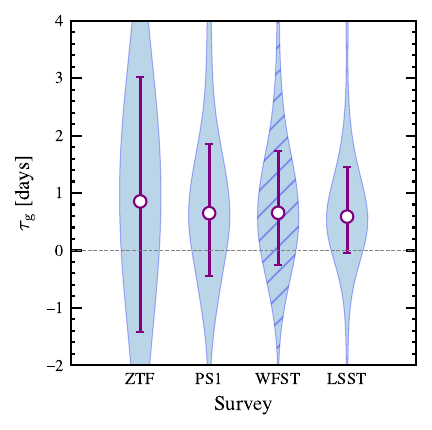}
 \caption{Same as Figure~\ref{fig:strategy_all}, but comparing the performance of different survey strategies of the ongoing and upcoming time domain surveys (see Table~\ref{tab:survey}), adopting the same $\mathcal{Y} = 1$ and $\mathcal{M} = 6$.}
\label{fig:agn_average_survey}
\end{figure}

\begin{deluxetable*}{ccccccl}\floattable
\caption{Information of four main time domain surveys\label{tab:survey}}
\tablewidth{0.8pt}
\tablehead{
\colhead{Survey} &
\colhead{Filter} &
\colhead{$r$-band 5$\sigma$ detection limit}& 
\colhead{Cadence} &
\colhead{Coverage} &
\colhead{SNR$_{\sigma}$ \tablenotemark{d}} &
\colhead{Advantages of WFST \tablenotemark{e}}\\
    & & (mag) & (day) & (deg$^2$) & 
}
\startdata
LSST DDF \tablenotemark{a} & $ugriz$ & $24.7$ &  2 & 40 & 9 & I, II \\ 
WFST DHS & $(u)gri(z)$ & $23.7$  &1 & 720 & 3  & I. finer cadence, \\
& & & & & & II. larger sky region, \\
& & & & & & III. deeper in detection limit, \\
& & & & & & IV. more quasi-simultaneous bands. \\
PS1 MDS \tablenotemark{b} & $gr(iz)$ & $23.3$ & 3  & 70 & 3 & I, II, III, IV \\ 
ZTF $3\pi$ \tablenotemark{c} & $gr(i)$ & $20.6$ & 3 & $\sim 30,000$ & 3 & I, III, IV \\
\enddata
\tablenotetext{a}{
The survey strategy of the LSST DDF has not yet been settled down. Hence, we adopt that proposed by \cite{Brandt2018}: 4 visits in $u$, 1 visit in $g$, 1 visit in $r$, 3 visits in $i$, 5 visits in $z$ for every two nights. The $r$-band 5$\sigma$ detection limit and the coverage of the LSST DDF are also taken from \cite{Brandt2018}.}
\tablenotetext{b}{
According to \citet{Chambers2016}, the Medium Deep Survey (MDS) of the PS1 repeats monitoring $griz$ bands in a 3-day cycle: $gr$ in the first night, $r$ in the second, and $z$ in the third. The $r$-band 5$\sigma$ detection limit and the coverage of the PS1 MDS are also taken from \cite{Chambers2016}.}
\tablenotetext{c}{ZTF maps a $3\pi$ sky at cadences of $\sim 3$ days in $gr$ bands mainly \citep{Graham2019}. The $r$-band 5$\sigma$ detection limit is taken from \citet{Masci2019}.}
\tablenotetext{d}{
The SNR$_{\sigma}$ equals $\langle \sigma_{\rm rms} \rangle / \sigma_{\rm e}$, where $\langle \sigma_{\rm rms} \rangle$ we simulate is 0.083~mag in the case of $\mathcal{Y}=1$ and $\mathcal{M}=6$.
A typical $\sigma_{\rm e}$ of $\simeq 0.03$~mag (or SNR$_{\sigma} \simeq 3$) is adopted for ZTF \citep[e.g.,][]{Jha2022, guo2022,GuoH2022}, PS1 \citep{Suberlak2021}, and WFST, while a 3 times better is assumed for LSST.}
\tablenotetext{e}{
Four potential advantages of the WFST DHS are nominated and the corresponding points are itemized when compared with the relevant surveys.
}
\end{deluxetable*}

Besides WFST, there are many ongoing and upcoming time domain surveys, such as PS1, ZTF, and LSST. Exploring the inter-band variation properties of AGNs is usually itemized as one of their scientific goals.
Table~\ref{tab:survey} presents the general information of these surveys compared to the WFST DHS. Figure~\ref{fig:agn_average_survey} illustrates the resultant performance of these surveys on the lag measurement, adopting the same $\mathcal{Y} = 1$ and $\mathcal{M} = 6$ for N5548-like AGNs.

Compared to ZTF, WFST has a better performance on the lag measurement thanks to its advantage of quasi-simultaneous observations in the $gri$ bands with a finer cadence of approximately 1 day.

The performances between PS1 and WFST are comparable. Although the PS1 has a sparser cadence than the WFST, the inclusion of the PS1 $z$ band indeed helps improve the quality of the lag measurement. Instead, the WFST DHS has a $\simeq 10$ times larger sky coverage and thus could give rise to a higher accuracy on the lag measurement once ensemble averaging the lag-wavelength relations of a larger sample of AGNs.

The performance of the LSST DDF is appealing. Nevertheless, the WFST DHS, with a daily cadence in $gri$ bands and a large sky coverage, makes it a competitive survey for exploring the inter-band lag of AGNs.

\section{Conclusions}\label{sect:conclusion}

In this work, we analyze the random behavior of the lag-wavelength relations based on two models for AGN variability, i.e., the thermal fluctuation model and the reprocessing model with DCE. Our simulations reveal that for both models the measured lags can differ from one finite duration/baseline to another, owing to the random nature of AGN variability. 

Given a finite duration/baseline, the thermal fluctuation model predicts a larger scatter in the lags, more violent lag evolution, and fewer cases of the $u$-band excess than the reprocessing model with DCE. 
In addition, a positive correlation between the inter-band lag and the variation amplitude is expected by the thermal fluctuation scenario, but not the reprocessing model with DCE. 
Future long-term, high-cadence monitoring on a sample of AGNs will be essential to test these predictions and refine models.

We suggest that for both models averaging the lags measured in repeated and non-overlapping baselines can achieve a stable lag. The longer the duration/baseline, the larger the averaged lag. 
The averaged lag would get saturation for a sufficiently long LC. 
Obtaining the average lag for an individual AGN requiring a sufficiently long LC is observationally challenging. 
Instead, averaging the lags of a sample of AGNs with similar physical properties is achievable and can be equivalently used to constrain model parameters. 

Finally, we perform a series of simulations to assess the observational effects of diverse conditions on the lag measurement based on the thermal fluctuation model. We suggest an optimal strategy, that is, a daily cadence in $gri$ bands with 6 months of continuous monitoring per year, for studying the inter-band lag of AGN variability with the upcoming formal WFST DHS. The resultant WFST measurements on the inter-band lags are expected to be better than ZTF/PS1 and competitive with LSST.

\section*{Acknowledgement}
We are grateful to the anonymous referee for many constructive suggestions, which have significantly improved our manuscript, and Xu Kong for helpful comments. This work is supported by the National Key R\&D Program of China (grant No. 2023YFA1608100) and the National Science Foundation of China (grant Nos. 12373016, 12033006, 12025303, 12173037, and 12233008). 
Z.Y.C. is supported by the science research grants from the China Manned Space Project under grant no. CMS-CSST-2021-A06 and the Cyrus Chun Ying Tang Foundations.
L.L.F. acknowledges the CAS Project for Young Scientists in Basic Research (No. YSBR-092) and the Fundamental Research Funds for the Central Universities (WK3440000006). 
Z.Z.H. acknowledges the National Postdoctoral Grant Program (No GZC20232990).
Z.C.H. is supported by the National Natural Science Foundation of China (nos. 12222304, 12192220, and 12192221).
G.L.L acknowledges the support from the China Manned Space Project (the second-stage CSST science project: “Investigation of small-scale structures in galaxies and forecasting of observations”, and the National Natural Science Foundation of China (no. 12273036).

\bibliographystyle{aasjournal}
\bibliography{ms}
\end{document}